\documentstyle[12pt]{article}
\advance\textheight by \topskip
\begin{document}
 \begin{flushright}
SU-ITP-96-56\\
hep-th/9612004\\
December 1, 1996\\
\end{flushright}
\vspace{1 cm}
\begin{center}
\baselineskip=16pt

{\Large\bf    WRAPPED  SUPERMEMBRANE}

\vskip 2cm

{\bf Renata
Kallosh}\footnote { E-mail:
kallosh@physics.stanford.edu}\\
 \vskip 0.8cm
Physics Department, Stanford University, Stanford,   CA 94305-4060, USA\\
\vskip .6cm

\vskip 1 cm

\end{center}

\vskip 1 cm
\centerline{\bf ABSTRACT}
\begin{quotation}

The Hamiltonian of the wrapping and KK modes of the supermembrane is identified
with the
SL(2, Z) symmetric axion-dilaton black hole  mass formula. It means that the
supermembrane  with KK modes   wrapped $m$ times around the space-time torus of
compactified dimensions induces the superpotential breaking spontaneously N=2
down to N=1 SUSY.  The supersymmetry breaking parameter Lambda is inversely
proportional to the area A of the space-time torus around which the
supermebrane wraps. The restoration of supersymmetry as well as
decompactification of  higher dimensions of  space-time are forbidden by the
requirement of stability of the wrapped supermembrane. This may be the ultimate
reason why supersymmetry is broken in the real world.

\end{quotation}

\newpage

The Bekenstein-Hawking entropy of many dilatonic black holes have been
understood from the microscopic string theory. The axion-dilaton black holes
with manifest $SL(2, {\bf Z})$ symmetry \cite{KO} seem to be more difficult to
understand. The interest to such black holes is related in particular to the
fact their  mass formula has a mysterious appearance \cite{KSUP} in the
potential \cite{N2break} which generate the spontaneous breaking of
$N=2\rightarrow N=1$ SUSY. It was suggested in \cite{KSUP} that
Fayet--Iliopoulos terms
  can be considered proportional to the electric and magnetic charges of the
dyonic black holes. Upon such identification  the potential was found to be
proportional to the square of the black hole mass.
The fixed values of the
moduli near the black hole horizon \cite{FKS} were found to correspond exactly
to the minimum of this
potential, as predicted in the general case \cite{FK}.  And finally, the value
of the  moduli-independent part of the potential at the minimum was found to be
 proportional to the axion-dilaton
black hole entropy.

Having established the relation between the potential  and the black holes was
interesting but  the puzzle remained as to why those black holes should be
around and how they may generate the potential \cite{KL}.

The issue of stability of extremal black holes \cite{FKS,FK} aquires a new
perspective when applied to the study  of the supermembrane \cite{BST}
Hamiltonian \cite{ham}.
The basic fact about the supermembrane is its instability \cite{WLN} unless it
is wrapped around some compactified dimensions \cite{Duff}. This is the clue to
the solutions of the problem. Whereas we do not have any particular reason to
think that the black holes with non-vanishing electric and magnetic charges
have to be present, we may conjecture that   a supermembrane is a fundamental
object which exists. If it exists it better be stable! This  requires the
non-vanishing wrapping number of the supermembrane $m$.

The new mechanism of partial SUSY breaking \cite{N2break} requires the
potential of the form
 \begin{equation}
V=  \Lambda^4  \left [{|e  +\xi \tau|^2 + m^2 \over \tau_2 }\right] \ ,
\label{APT}\end{equation}
where $\Lambda$ is the supersymmetry breaking scale related to the gravitino
mass as $m_{3/2}\sim {\Lambda^2 \over M_{\rm Pl}}$.
The values of the gauge couplings which minimize the potential are
of the form
\begin{equation}
\tau_{\rm min} = (\tau_1 + i \tau_2)_{\rm min} = {- e + i |\xi| \over m}\ .
\label{mod}
\end{equation}
The mass which the complex scalars acquire is proportional to
 $ m |\tau|_{\rm min}$. Also one combination of the two original fermions,
present in N=2 theory acquires the same mass, whereas the second combination of
fermions remains massless. In this way one  gets  from one massless vector
multiplet of N=2 SUSY one massless vector multiplet of N=1 SUSY and one massive
chiral multiplet of N=1 SUSY  with the mass proportional to
$ \sqrt {e^2 + \xi^2 }$. In this way
 N=2 supersymmetry is broken down to N=1 spontaneously.

The electric FI terms  $ (e ,  \xi )$ without the magnetic $m$ obviously would
not work! Thus we may attribute the stability of the state with  spontaneous
breaking of  $N=2\rightarrow N=1$ supersymmetry to the actual stability of the
object: wrapped supermembrane.

Our goal is to identify the  ADM mass formula of the axion-dilaton black hole
\cite{KO} with the Hamiltonian of the  supermembrane, wrapped around the torus.
The mass formula for this black hole was related to the SUSY breaking potential
before \cite{KSUP}. It required the identification of the electric and magnetic
black hole charges with the parameters $(e, \xi, m)$ in the potential.
One finds that the part of the supermembrane Hamiltonian describing the
wrapping and KK modes is given by
\begin{equation}
  H_0 = H_{\rm KK} + H_{\rm wr} = { 4\pi^2\over A } \left ({|l_1  - l_2 \tau|^2
+ m^2 \over \tau_2 }\right)\ ,
\label{Ham}\end{equation}
where $l_1, l_2, m$ are positive integers.
To get this form of the Hamiltonian one has to use the following relation:
\begin{equation}
 A^3   ={1\over \tau_2}\left ( {2\pi\over T }\right)^2 \ .
\end{equation}
Here $T$ is the tension of the supermembrane and $A$ is the area of the torus
around which the membrane is wrapped,  $\tau=\tau_1 + i \tau_2 $ is the modular
parameter of a torus. Thus for our purpose of identifying the axion-dilaton
black hole mass formula with the state of  the supermembrane it is important to
relate the area of the torus, its modular parameter and the  tension.
The  numbers $l_1, l_2$
 are Kaluza-Klein modes of the membrane on a torus and $m$ is the wrapping
number.
To accomplish the identification of the supermembrane Hamiltonian with the
supersymmetry breaking potential $ V \sim H_0$ we have to identify
\begin{equation}
l_1=-e, \qquad l_2 =\xi \ .
\end{equation}
Finally, the supersymmetry breaking parameter is proportional to the inverse of
the area
of the wrapped supermembrane:
\begin{equation}
\Lambda^4 \sim {1\over A} \ .
\end{equation}
The restoration of extended supersymmetry requires the decompactification of
the compactified dimensions
\begin{equation}
A\longmapsto \infty\ ,  \qquad \Lambda\longmapsto 0 \ .
\end{equation}
In such a process the supermebrane state is going to be destabilized. For this
not to happen, the extended supersymmetry has to be permanently broken! Thus,
spontaneous breaking of supersymmetry is the condition for the existence of a
stable wrapped supermembrane.

The Hamiltonian \cite{ham}  has the structure of the axion-dilaton black hole
mass formula \cite{KO}. This means that we may apply the extremization
procedure \cite{FK} to it. The proper procedure will be to vary the modular
parameter of the torus while keeping fixed its area $A$ and the KK modes and
the wrapping number,
\begin{equation}
{\partial H_O \over \partial \tau}|_{A, l_1, l_2 ,m} = 0 \ .
\end{equation}
The value of the modular parameter  which minimizes the supermembrane
Hamiltonian $H_0$ is exactly given in eq. (\ref{mod}). The minimal value of the
Hamiltonian, dilaton and axion  are
\begin{equation}
  (H_0)_{\rm min} =  { 8\pi^2\over A }  l_2 m \qquad (\tau_i )_{\rm min} = {l_i
\over m} \qquad i=1,2 \ .
\label{minH}\end{equation}

In what follows we will present few details on the derivation of the
Hamiltonian  (\ref{Ham}). We are using here the construction developed in
\cite{S,Russo}.
 A single valued wave function is given in terms of the coordinates on the
torus,
\begin{equation}
\phi_{l_1,l_2} = \exp { {i\over 2\pi \tau_2} [ z(l_1- l_2 \bar \tau) + \bar
z(l_1- l_2   \tau)]  }\ .
\end{equation}
The contribution to the mass formula is
\begin{equation}
- {\tau_2\over 4\pi^2  } \partial_z \partial_ {\bar z} \;  \phi_{e,\xi} =
{|l_1- l_2  \tau|^2  \over  \tau_2  } \; \phi_{l_1,l_2}\ ,
\end{equation}
which explains the origin of the first term in (\ref{Ham}):
\begin{equation}
H_{\rm KK}= { 4\pi^2\over A } \left({|l_1  - l_2 \tau|^2  \over \tau_2 }\right)
\ .
\end{equation}
The origin of the $m$ term is the following. The light-cone gauge Hamiltonian
of the supermembrane \cite{ham} has a term
\begin{equation}
H = \pi^2 T^2 \int d\sigma d\rho \{ X^{11}, X^{10}\}^2\ .
\label{Poisson}\end{equation}
The contribution to this term comes from  the space-time coordinates which are
linear functions of the membrane coordinates. The torus of the space-time can
be mapped into the torus of the world-volume. In the simplest case of the
rectangular torus  this mappings was described in \cite{Russo},
\begin{eqnarray}
X^{11}&=& \omega_1 R_{11} \sigma + \omega_2 R_{10} \rho \nonumber\\
 X^{10}&=& \upsilon_1 R_{11} \sigma +\upsilon_2 R_{10} \rho \ .
\end{eqnarray}
The wrapping number that counts how many times the toroidal membrane is wrapped
around the target space-torus is given by
\begin{equation}
m= {1\over A}\int d\sigma d\rho \{ X^{11}, X^{12}\} = \omega_1 \upsilon_2 -
\omega_2 \upsilon_1\ .
\end{equation}
The contribution of the wrapped states to the part of the Hamiltonian of the
supermebrane (\ref{Poisson}) is given by
\begin{equation}
H_{\rm wr} = \pi^2 T^2 \int d\sigma d\rho \{ X^{11}, X^{12}\}^2 = {1\over 2}
A^2  T^2  m^2 \ .
\label{hamwrap}\end{equation}
In a more general case one has to choose a canonical basis of homology cycles
$A,\; B$ and $C,\; D$  for the  space-time torus  and for the membrane torus,
respectively. For the space-time torus we have
\begin{equation}
\oint_{A}{\omega(z)dz}=1, \qquad \oint_{B}{\omega(z)dz}=\tau \ ,
\end{equation}
where $\omega(z)dz$ is Abelian differential on the torus.
The homology cycles of two toruses are related (``homologically equivalent")
\begin{eqnarray}
C&=&\omega_1 A +  \omega_2 B \nonumber\\
D&=&\upsilon_1 A +  \upsilon_2 B \ .
\end{eqnarray}
The integers $\omega_1,\; \omega_2$, called winding numbers, count how many
times the cycle $C$ wraps around the cycles $A$ and $B$. The integers
$\upsilon_1,\; \upsilon_2$ count how many times the cycle $D$ wraps around the
cycles $A$ and $B$.
Under the $SL(2, {\bf Z})$ symmetry the modular parameter of the space-time
torus $\tau$
transforms as
\begin{equation}
\tau' = {d\tau - b \over c\tau + a} \ .
\end{equation}
The homology basis also changes
\begin{equation}
\left (\matrix{
A'\cr
B'\cr
}\right )= \pmatrix{
d & -b \cr
-c & a \cr
} \left (\matrix{
A\cr
B\cr
}\right ) \ .
\end{equation}
In terms of the transformed basis we will have
\begin{eqnarray}
C&=&\omega_1' A' +  \omega_2 B' \nonumber\\
D&=&\upsilon_1' A '+  \upsilon_2' B' \ .
\end{eqnarray}
The new winding numbers are related to the old ones as follows
\begin{equation}
\left (\matrix{
\omega_2'\cr
\omega_1'\cr
}\right )= \pmatrix{
a & b \cr
c & d \cr
} \left (\matrix{
\omega_2\cr
\omega_1\cr
}\right ) \ , \qquad \left (\matrix{
\upsilon_2'\cr
\upsilon_1'\cr
}\right )= \pmatrix{
a & b \cr
c & d \cr
} \left (\matrix{
\upsilon_2\cr
\upsilon_1\cr
}\right ) \ .
\end{equation}
Now the definition of the membrane wrapping number is clear: it is an  $SL(2,
{\bf Z})$ invariant exterior product of two duality vectors, winding vectors
$(\omega_1, \omega_2)$ and $(\upsilon_1, \upsilon_2)$:
\begin{equation}
m=(\omega_2, \omega_1) \Omega \left (\matrix{
\upsilon_2\cr
\upsilon_1\cr
}\right ) = (\omega_2, \omega_1) \pmatrix{
0 & -1 \cr
1 & 0 \cr
} \left (\matrix{
\upsilon_2\cr
\upsilon_1\cr
}\right ) = \omega_1 \upsilon_2 - \omega_2 \upsilon_1 \ .
\end{equation}
This clarifies the meaning of the supermembrane wrapping number $m$. The
constant parts of the SUSY transformation of the fermions in the spontaneously
broken phase are \cite{N2break}
\begin{equation}
\delta {\chi + \lambda \over \sqrt 2} =0 \ , \qquad \delta  {\chi -
\lambda\over \sqrt 2}  = -2i m (\eta_1 - \eta_2) \ ,
\end{equation}
where the massless goldtsino ${\chi - \lambda \over \sqrt 2}$ has the
supermembrane wrapping number $m$ in the transformation law.

Thus we have learned that the non-vanishing  Fayet--Iliopoulos terms are
required for the topological stability of the supermembrane. Specifically the
magnetic FI term, which is of crucial importance for the new mechanism of
spontaneous breaking of extended supersymmetry, is proportional to the wrapping
number $m$ of the  supermembrane state.

The idea developed in this paper is as simple as in case of anomalies: first
one should find the theory which is generically ``bad" and try to make it
``good". If one succeeds in a non-trivial way, this may be the
desirable theory. In case of anomalies the mechanism was to produce the correct
set of the particles of the theory, and it worked, it explained and predicted
which particles should exist in nature. Still the origin of the Higgs terms in
the standard model potential $- \mu^2 \phi^2 +
\lambda \phi^4$ remains  obscure.

The mechanism of spontaneous breaking of extended supersymmetry described above
is based on the fact that  breaking of supersymmetry is the condition for the
existence of a stable wrapped supermembrane. If the extended supersymmetry
where restored, there would be no stable state of a supermembrane. Again, as in
case with  anomalies we were lucky to have a ``bad" theory which hopefully can
be made ``good".
The  necessary condition for this is the existence of the wrappings of the
supermembrane simultaneosly with the
spontaneous breaking of extended supersymmetry and existence of compactified
dimensions of the space-time around which the supermembrane wraps.

This gives a serious support to the idea that wrapped supermembrane is
fundamental. The stability of the total configuration seems to be the ultimate
reason for the supersymmetry in the real world to be broken
and higher dimensions  compactified.

\vskip 0.5 cm

 Stimulating discussions with   A. Linde, J. Rahmfeld and S.-J. Rey   are
gratefully
acknowledged. The work  is supported by the  NSF grant PHY-9219345.

\end{document}